# Development of a Load Profile Generator for Non-road Mobile Machinery

Serhiy Kapustyan [a,†], Pranav Tetey [b,†], Thomas Grube[b, *], Jochen Linssen[b]

[a] Forschungszentrum Jülich GmbH, Institute of Climate and Energy Systems – Juelich Systems Analysis (ICE-2), 52425 Jülich, Germany

[b] Technische Hochschule Mittelhessen, CCCE, Faculty of Information Technology-Electrical Engineering - Mechatronics, 61169 Friedberg, Germany

[†] Equally contributed

* Corresponding author: Thomas Grube, Forschungszentrum Jülich GmbH, Institute of Climate and Energy Systems – Juelich Systems Analysis (ICE-2), 52425 Jülich, Germany

## Abstract

This research presents a Load Profile Generator model for non-road mobile machinery, which depicts the most common operational profiles that reflect real-world conditions. This technological bottom-up model enables users to parameterize specific machines for simulation and observe their power demand at the actuator interfaces. The application of the Load Profile Generator covers different non-road mobile machinery categories, such as construction, agriculture, industrial and forestry. In this study, the Load Profile Generator was used to demonstrate common operations of material handler and forest forwarder, which have been validated against real-world data to match the results. The usage of Load Profile Generator aids engineers in evaluation machines performance by developing operation strategies. It opens doors to further systems analysis as it can serve as an interface for energy demand calculations. The model's results are accurate enough to provide a sufficient understanding of a wide range of non-road mobile machinery load profiles as well as insights about alternative fuel and power supply concepts helping in optimising the system simulation.

## Introduction

In recent years, the automotive sector has become increasingly concerned about fuel consumption, emissions and sustainable operations. Understanding the operational behavior of the vehicles is important for analyzing their efficiency and fuel consumption, and for finding ways to reduce their environmental impact [1].

According to Hagan et al. Non-Road Mobile Machinery (NRMM) is defined as a vehicle that may or may not include a carriage and wheels, and that is not used to carry people or goods on the road [2]. There are several categories of NRMM, although challenges for a consistent categorization exist. Construction machinery refers to self-propelled, heavy-duty vehicles designed to perform specific construction tasks, such as excavators and loaders [3,4]. Agricultural machinery refers to mechanical tools and machinery used in farming activities [5]. Forestry machinery, for example, includes modern forwarders and skidders that perform various functions [6,7]. A common example of mobile industrial machinery is the forklift truck [8].

To better understand the distribution and impact of NRMM across different sectors, the fuel consumption patterns and machine registration data in Germany are examined. Germany was chosen due to a good availability of data and the similarity that may exist to a great number of other countries, e.g., in Europe or North America. Figure 1 shows that construction and agricultural machines account for more than 95% of the non-road mobile machinery sector [8]. Therefore NRMM accounts for 49 TWh (177 PJ) of fuel consumption, which represents about 9% of the German transport sector's fuel demand of 538 TWh (1936 PJ) in 2022 [9]. The distribution of NRMM fuel consumption was taken from the German federal state of Baden-Württemberg [8,10] .

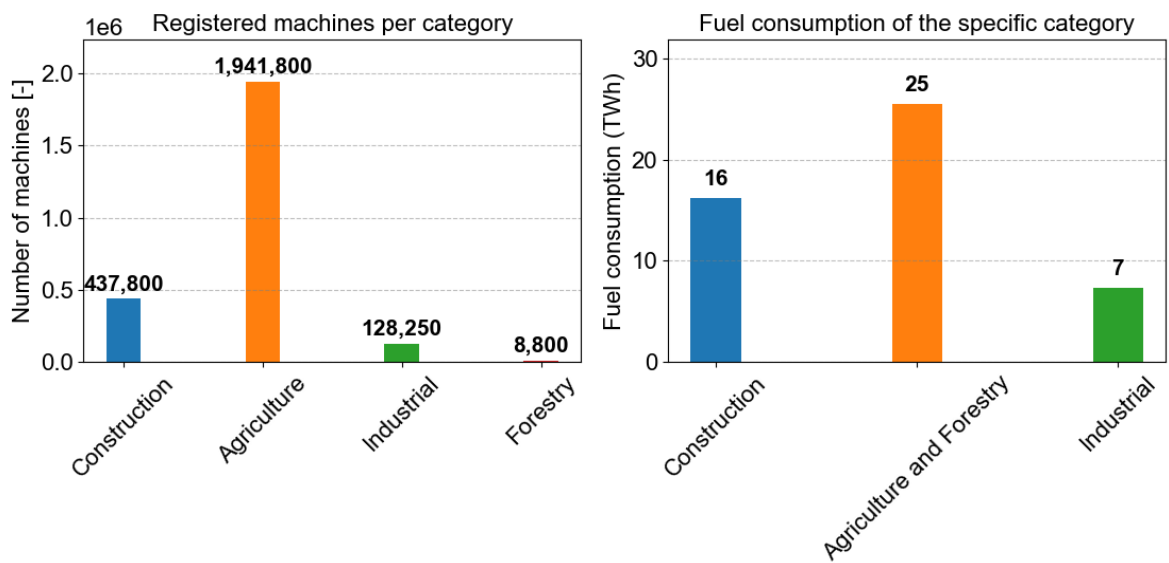

**Figure 1:** Machinery distribution and fuel consumption of NRMM per category in Germany [8,10]**.**

In Germany, within the category of mobile sources, non-road mobile machinery accounts foris responsible for about 45% of particulate matter emissions, while its share in nitrogen oxides, carbon monoxide, and hydrocarbon emissions is between 13% and 18% [11]. As alternative solutions are being developed, a systematic analysis is needed to evaluate the performance of NRMM. Due to the diversity of operational tasks, a method of generating machine-specific load profiles is necessary to cover the full range of a machine's behaviour.

Such load profiles refer to the combination of speeds, positions, forces, and power outputs associated with either individual machine components or the machine itself. Drawing on the complexity of such load profiles, this study presents a flexible load profile generator representing the operational behavior of a variety of NRMM, including the scaling of the machinery under consideration.

With many types of machines on the market, it makes sense to concentrate on the most relevant ones. Figure 2 highlights machine segments from four major categories selected for this study: construction, agriculture, forestry, and industry. These segments were chosen based on their contribution to total energy consumption and sales numbers from the a transport emission model TREMOD which covers motorised traffic in Germany including NRMM [8].

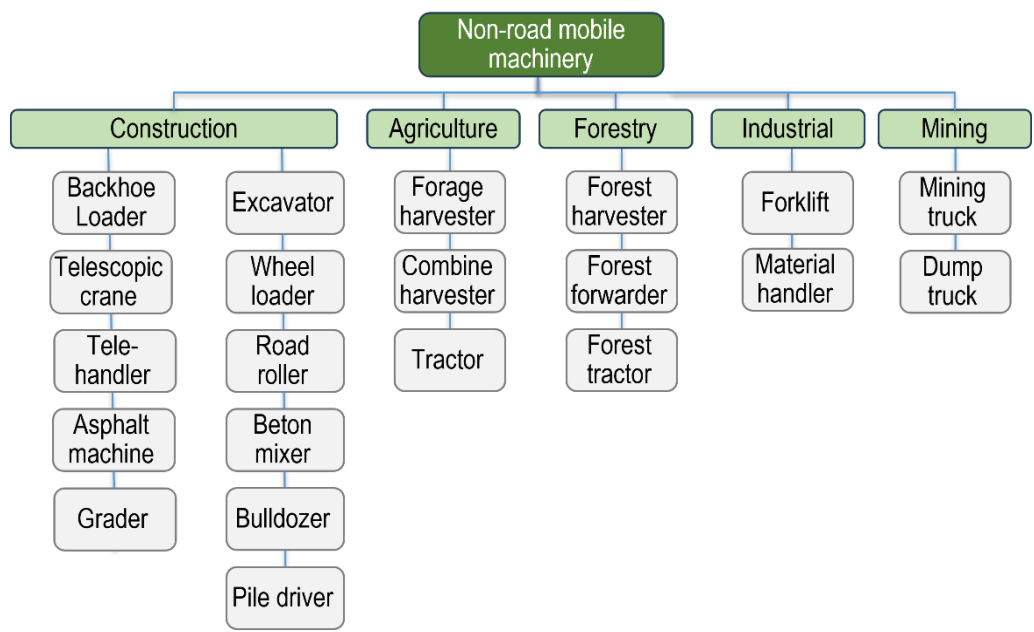

**Figure 2:** NRMM distribution diagram.

The scientific literature survey carried out in this study revealed that only a few studies investigated the modeling of load profiles for a variety of non-road mobile machinery. Fecke [12] developed a load profile generator that incorporates different work cycles derived from real-world tests for different excavator and wheel loader categories. Agostini et al. [13] evaluated the energy efficiency of the front hoe of a micro-excavator considering digging and leveling cycles. Wong et al. developed a tracked vehicle performance model for military applications [14], focusing on track and grading resistance for different terrain conditions. In addition, case studies of typical mining loader work cycles, taking into account different levels of driver experience, were presented by Luostarinen [15]. A multibody dynamics-based tractor model was also developed by Goswami et al. [16], using a co-simulation environment to evaluate performance under dynamic load conditions.

The studies mentioned above certainly contribute to a deeper understanding of NRMM performance. However, a comprehensive tool that integrates all relevant machine types in a scalable manner with their specific operating conditions was not presented, so far. To address the key challenges and explore the potential applications of the load profile generator, this work focuses on three research tasks.

- First, the operation tasks must be designed to represent the real conditions of non-road mobile machinery. For this purpose, the operating environments and constraints of the selected machines are analyzed.
- Second, bottom-up machine-specific mathematical models are to be developed in order to determine the force and power requirements based on different operating tasks. This includes the creation of functions to calculate forces and torques considering machine-specific geometry.
- Third, the power demand calculations must be validated or checked for plausibility against available data to ensure their accuracy.

As there is only little data publicly available, we selectively compare study results with normalized data from a real-world material handler operation and with data from hardware-in-the loop simulations related to a forest forwarder. Such a bottom-up model, representing the different power requirements of the most commonly used NRMMs in different sectors, supports system analysis from individual sites to larger regions, such as Germany. The results can serve a range of applications, including energy consumption calculations, infrastructure development planning and cost assessment.

## Methodology

This section outlines the methods used to develop the load profile generator for non-road mobile machinery. First, an overview of the methodology is presented, then the introduced algorithm is described in detail, and finally the bottom-up mechanical models of the machinery are presented, which allows for power calculation.

### Design of the load profile generator

Figure 3 illustrates the overall design of the load profile generator, consisting of a working task selector, time series generator, mechanical models of the machinery, and a load profile designer. First, the working task selector is used to chose the machine of interest and the tasks to be performed, as well as tasks sequences, velocities, positions, and respective boundary conditions. Following that, the time series generator uses the selected tasks to compute the kinematic profiles of position, velocity, and acceleration. Next, the mechanical models specific to each machine are used to calculate the resulting forces and power requirements. Finally, the load profile designer arranges the operational function outputs in the desired sequence while considering the specifics of each task. These calculations vary depending on the type of work the machine performs.

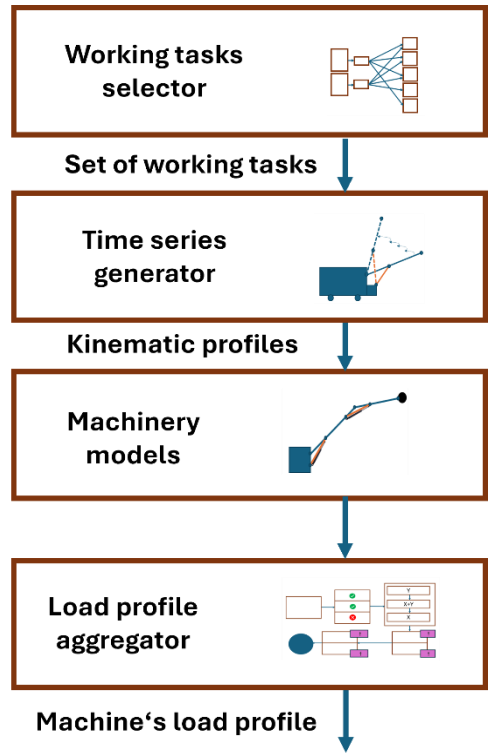

**Figure 3: Proposed design of the load profile generator for non-road mobile machinery.**

**Working task selector**

Figure 4 shows two selected machine classes each with one machine and associated operational functions, exemplifying respective combinations that form the basis of the working profile design. These two machines were chosen because they show comparatively complex behavior and belong to distinct categories of machinery. The distinguishing feature of this working task pool is that different machine types could make use of the same basic operational function. In addition, these operations can be combined and their parameters adjusted. Therefore, the operations of the machine under consideration is represented by a combination of working tasks.

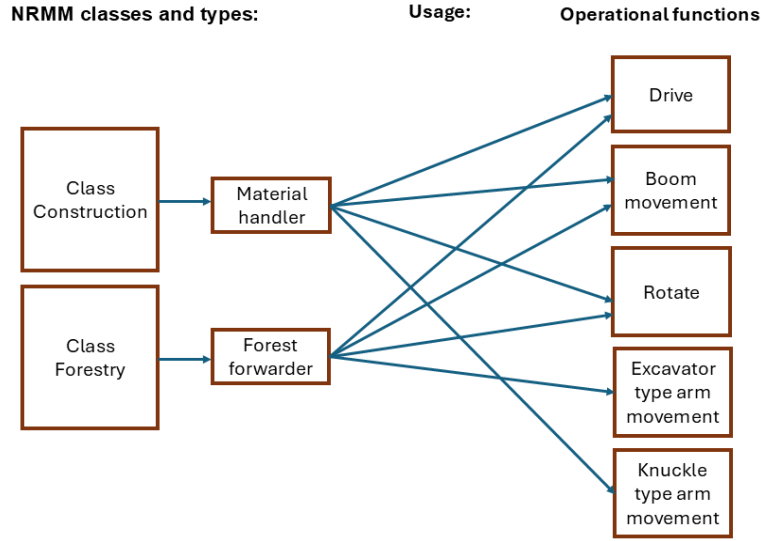

**Figure 4:** Machine classes and types and their operational functions. The items presented here deal with two machinery types: material handler and forest forwarder. Other classes and operational functions that relate to NRMMs defined in Figure 2 are available in the load profile generator but not considered here.

### Time series generator

The time series generator is used to create position, velocity, and acceleration time series for different machine operational functions. The formula for **position profile in discrete time** is derived by using the Mathematica software by considering the standard acceleration profile with initial and final position and maximum velocity as parameters by its solver [17]:

$$T[0] \leq t < T[1]: \quad v_{max} \cdot \frac{(t-T[0])^2}{2(T[1]-T[0])}$$

$$T[1] \leq t < T[2]: v_{max} \cdot \left(t - \frac{T[1]+T[0]}{2}\right) \quad Exp.\ 1$$

$$T[2] \leq t < T[3]: \quad v_{max} \cdot \left(\frac{-t^2/2 + tT[3]}{T[3]-T[2]} + \frac{(T[2]^2/2 - T[2]T[3])}{T[3]-T[2]} + \frac{T[2]-T[0]}{2} + \frac{T[2]-T[1]}{2}\right)$$

Where 'T[0]' is the start time, while 'T[1], T[2], T[3]' represent specific points in time for the transition of the profile between the phases of acceleration, constant movement and deceleration, 'v' denotes the speed. The movement of the machine is divided into distinct phases based on time intervals:

- T[0] ≤ t < T[1]: acceleration phase, where velocity stays linear in time;
- T[1] ≤ t < T[2]: constant speed phase, where quadratic term is absent as there is no acceleration;
- T[2] ≤ t < T[3]: deceleration phase, with negative quadratic term presenting the constant negative deceleration;

To ensure that task executions remain feasible, several measures must be implemented:

- *Problem case 1:* The end position is never reached at the specified speed.
- *Solution:* Calculate the required duration and adopt the increased duration.

$$s_{end,calc} < s_{end,set}: \quad t_{new} = f(\ s_{end,set}, a_{ramp,set}, v_{max,set})$$

Eq. 2

- *Problem case 2:* The end position is reached too quickly at the specified duration and maximum speed.
- *Solution:* Calculate the required maximum speed for this and adopt the reduced speed

$$t_{calc} < t_{set}: \quad v_{max,new} = f(\ s_{end,set}, a_{ramp,set}, t_{set})$$

Eq. 3

## Machinery models

Machinery models represent the mathematical functions of machinery components mechanics. These models represent a bottom-up calculation of the desired output of the power demand of different machinery components at specific interfaces, such as hydraulic cylinders and hydraulic motors, before entering the hydraulic transmission logic. For power demand calculations in this study, the maximum angular velocities and initial positions of the arm, boom, and cabin are used. To demonstrate its application, we decided to introduce two use cases. The first use case is the application of the load profile generator to a material handler, and the second is the application of the load profile generator to a forest forwarder.

### Example: Material handler

Material handlers are machines used in recycling, scrap and bulk material handling industries. They are characterized by their boom and arm system, which is typically longer than that of conventional excavators for the necessary reach and lift capacity. Material handlers are equipped with grabs and attachments like load hooks and scrap magnets.

### Boom

The center of mass of the boom is calculated firstly:

$$x_{boom} = \frac{L_{boom}}{2} \cos(\theta_1)$$

Eq. 4

The center of mass of the arm from the boom point *A*

$$x_{arm} = L_{boom} \cos(\theta_1) + \frac{L_{arm}}{2} \cos(\theta_1 + 180 - \theta_2)$$

Eq. 5

With bucket tip position

$$x_{bucket} = L_{boom} \cos(\theta_1) + L_{arm} \cos(\theta_1 + 180 - \theta_2)$$

Eq. 6

The arm system center of mass becomes

$$x_{arm,tot} = \frac{m_{arm} x_{arm} + m_{bucket} x_{bucket}}{m_{arm,tot}}$$

Eq. 7

Therefore, the overall center of mass are calculated as follows

$$x_{COM} = \frac{m_{boom} x_{boom} + m_{arm,tot} x_{arm,tot}}{m_{tot}}$$

Eq. 8

The torque and force calculations for the boom lifting are similar to those for an excavator, described as follows

$$M_{boom} = m_{tol} \cdot g \cdot x_{COM}$$

Eq. 9

The geometry for the cylinder mount can be calculated using the similarity from the publication of Nezhadali et al. [18]

$$x_c = \frac{z_c}{\sqrt{2}}$$

Eq. 10

$$y_c = -\frac{z_c}{\sqrt{2}}$$

Eq. 11

Where $z_c$ are the distance between points A and B, which are given depending on the boom length. The angle to horizontal line from the boom cylinder connection [18] is

$$\theta_c = \left( \frac{l_{joints(AB)} \cos(\theta_2) - x_c}{l_{joints(AB)} \sin(\theta_2) - y_c} \right)$$

Eq. 12

Finally, the force acting on the boom cylinder [19] is

$$F_{cyl,boom} = \frac{M_{boom}}{l_{joints(AB)} \cdot \sin(\theta_c - \theta_1)}$$

Eq. 13

The lowering boom force corresponds to

$$F_{cyl,boom} = p \cdot \frac{\pi}{2} \cdot (D_{cyl,boom} - D_{rod,boom})^2$$

Eq. 14

The boom can be brought down by releasing pressure from the pressure valves slowly. Therefore, only the atmospheric pressure $p$ must be provided in the rod cylinder chamber by the pump. $D_{cyl}$ represents the diameter of the boom cylinder, $D_{rod}$ the diameter of the boom rod, $l_{joints(AC)}$ is the distance between cylinder and the boom attachment on the upper carriage. The power needed to lift the boom with its cylinder is depending on the linear velocity of the cylinder $v_{cyl,boom}$

$$P_{cyl,boom} = F_{cyl,boom} \cdot v_{cyl,boom}$$

Eq. 15

Whereas the $v_{cyl,boom}$ depends on the change in the boom cylinder length

$$l_{cylinder(BC)} = \sqrt{(l_{joints(AB)} \cdot cos(\theta_1) - x_c)^2 + (l_{joints(AB)} \cdot sin(\theta_1) - y_c)^2}$$

Eq. 16

Figure 5 shows the corresponding arrangement of the boom forces.

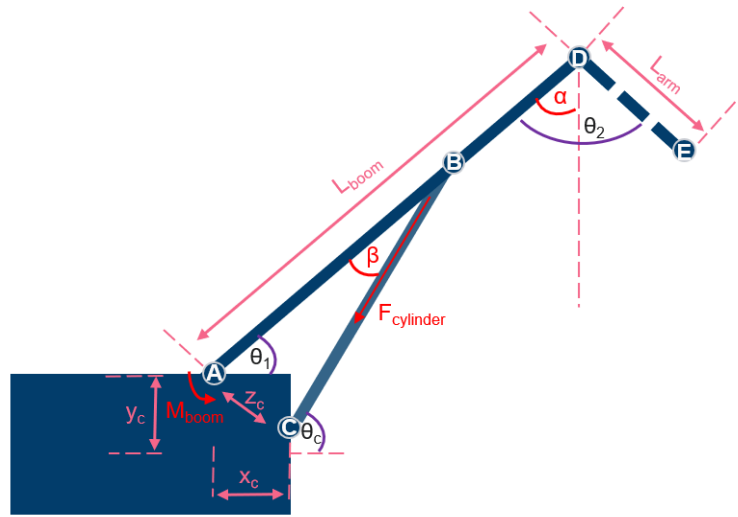

**Figure 5: Scheme of boom forces directions.**

**Arm**

The force for the arm of material handler is calculated differently from the arm of an excavator due to the position of the arm cylinder. The lifting torque, angle $\alpha$ and subsequently the force of the arm is calculated as follows

$$M_{Arm} = m_{grapple + load} \cdot g \cdot l_{Arm} \cdot sin(\alpha) + m_{Arm} \cdot g \cdot \frac{l_{Arm}}{2} \cdot sin(\alpha)$$

Eq. 17

$$\alpha = -90 + \theta_2 + \theta_1$$

Eq. 18

$$F_{cyl,arm} = \frac{M_{Arm}}{l_{joints(DC)} \cdot sin(\beta)}$$

Eq. 19

Where the cylinder length BC and the angle $\beta$ are represented through following formulas

$$\beta = \cos^{-1}\left[\frac{(l_{cylinder(BC)}^2 + l_{joints(DC)}^2 - l_{joints(BD)}^2)}{2 \cdot l_{cylinder(BC)} \cdot l_{joints(DC)}}\right]$$

Eq. 20

$$l_{cylinder(BC)} = \sqrt{l_{joints(BD)}^2 + l_{joints(DC)}^2 - 2 * l_{joints(BD)} \cdot l_{joints(DC)} \cdot \cos(\theta_1)}$$

Eq. 21

Where the lowering force is the similar to the boom one

$$F_{cyl,arm} = p \cdot \frac{\pi}{2} \cdot (D_{cyl,arm} - D_{rod,arm})^2$$

Eq. 22

and the power demand is determined by the linear velocity of the cylinder $v_{cyl,arm}$ given by the change in the cylinder length

$$P_{cyl,arm} = F_{cyl,arm} \cdot v_{cyl,arm}$$

Eq. 23

where $M_{Arm}$ is the moment acting on the arm, $l_{joint}$ is the length of the joint and $D_A$ is the arm cylinder diameter. Figure 6 presents the layout of the arm forces accordingly.

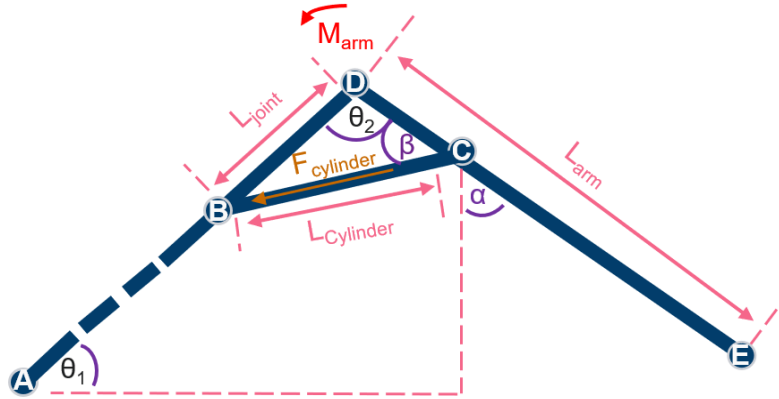

**Figure 6: Scheme of material handler arm forces directions.**

**Upper carriage swing**

Mounted on a platform above the undercarriage, the upper carriage allows the cabin to rotate up to 360 degrees. This is supported by bearings and powered by a hydraulic motor. The cab is additionally equipped with a counterweight at the rear to balance the weight of the boom, arm and load. To calculate the power requirement of the slewing operation, the moment of inertia of the rectangular body can be determined using the formula [20]

$$I_{cabin} = m_{cabin} \cdot \frac{(l_{cabin}^2 + w_{cabin}^2)}{12}$$

Eq. 24

By using the parallel axis theorem, the moment of inertia of the boom and arm arrangement can be calculated as [21,22]

$$I_{boom} = m_{boom} \cdot \frac{l_{boom}^2}{3}$$

Eq. 25

$$I_{arm} = m_{arm} \cdot \frac{l_{arm}^2}{12} + m_{arm} \cdot R^2$$

Eq. 26

where $d$ is the distance between the center of mass of the arm and the axis of rotation. It is equal to the center of mass of the boom and arm manipulator

$$R = \sqrt{x_{COM,arm}^2 + y_{COM,arm}^2}$$

Eq. 27

where $m$, $l$ are the mass and the length as referred to the figure, and $\theta_2$, $\theta_1$ are the angles of the boom and the arm presented in Figure 7. The total moment of inertia can be calculated as [23]

$$I_{total} = I_{cabin} + I_{boom} + I_{arm}$$

Eq. 28

The torque required to swing the cabin is given by

$$M = I \cdot \frac{d\omega_{rot}}{dt}$$

Eq. 29

and the power demand is determined by

$$P = M \cdot \omega_{rot}$$

Eq. 30

where $\omega$ is the angular velocity of the swing. Additionally the friction from the rotational bearings were taken from Liebherr calculations [24].

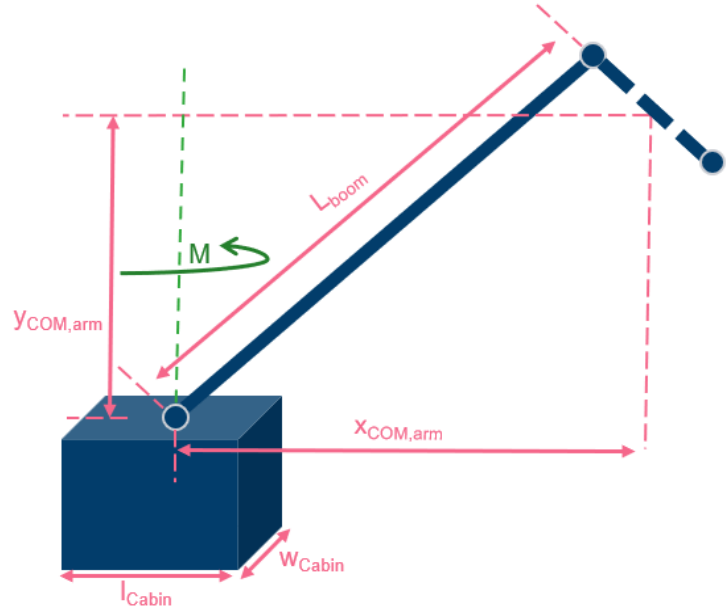

**Figure 7: Scheme of cabin with manipulator general swing modelling.**

**Scaling of the components**

Because force calculations for different machine sizes require specific parameters that are not publicly available, it is decided to obtain these parameters by scaling. This is done using the mass and parameters of the specific machine as a reference. For example, the **scaling factor** for the cabin mass is calculated as follows:

$$Scaling\ Factor_{ref} = \frac{m_{cabin,ref}}{m_{vehicle,ref}}$$

Eq. 31

This factor can then be applied to parametrize the newly configured machinery. For example:

$$m_{cabin,new} = Scaling\ Factor_{ref} \cdot m_{vehicle,new}$$

Eq. 32

Similar scaling formulas are used to determine other parameters, such as the length of machinery components.

### *Example: Forest Forwarder*

A forest forwarder is a vehicle used to transport cut trees at the forest site. It can drive on a range of terrains that are common in wooded regions. It has an attachment at the end of its arm which is used to gather the logs and put them into a separate compartment.

**Arm**

Firstly, the length of the cylinder of the arm is computed with the following formula (cf. Figure 8):

$$l_{cylinder(BC)} = \sqrt{l_{joint(CD)}^2 + l_{joint(BC)}^2 - 2 * l_{joint(CD)} \cdot l_{joint(BC)} \cdot \cos(\beta)}$$

Eq. 33

With

$$\beta = 180 - \theta_2$$

Eq. 34

The formula for the torque at joint point while lifting is given as follows

$$M_{arm} = W_{bucket+load} \cdot l_{arm+bucket} \cdot \sin(\alpha) + W_{arm} \cdot \frac{l_{arm}}{2} \cdot \sin(\alpha)$$

Eq. 35

$$\alpha = -90 + \theta_1 + \theta_2$$

Eq. 36

The force acting on the arm cylinder while lifting action is calculated as

$$F_{cyl} = \frac{M_{arm}}{l_{joints(CD)} * \sin(\beta)}$$

Eq. 37

During the lowering of the arm, the force is calculated in the same way as for the material handler.

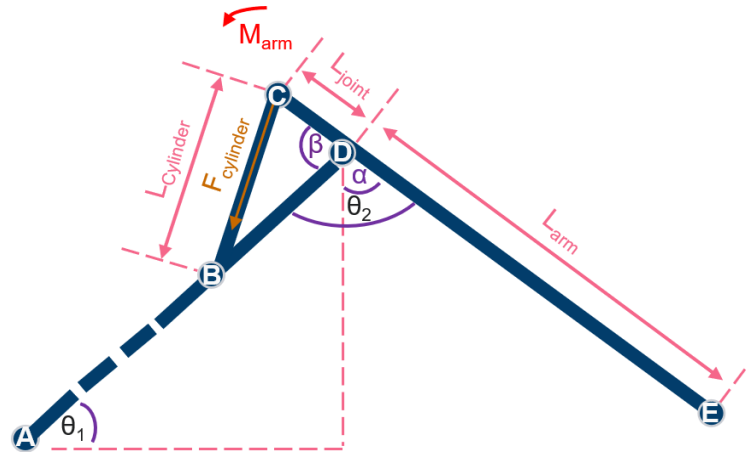

**Figure 8: Scheme of forest forwarder arm forces directions.**

## Load profile aggregator

The basic concept of the load profile aggregator is shown in Figure 9. Starting with machine selection, operational functions are then chosen and configured to retrieve a logical sequence of machine operations. The aggregator then outputs the desired load profile for the specified sequence of operations. A machine like an excavator can have two drive options, wheels or tracks, as can a material handler. Similarly, a forest forwarder can have all the functionality of an excavator except for the bucket.

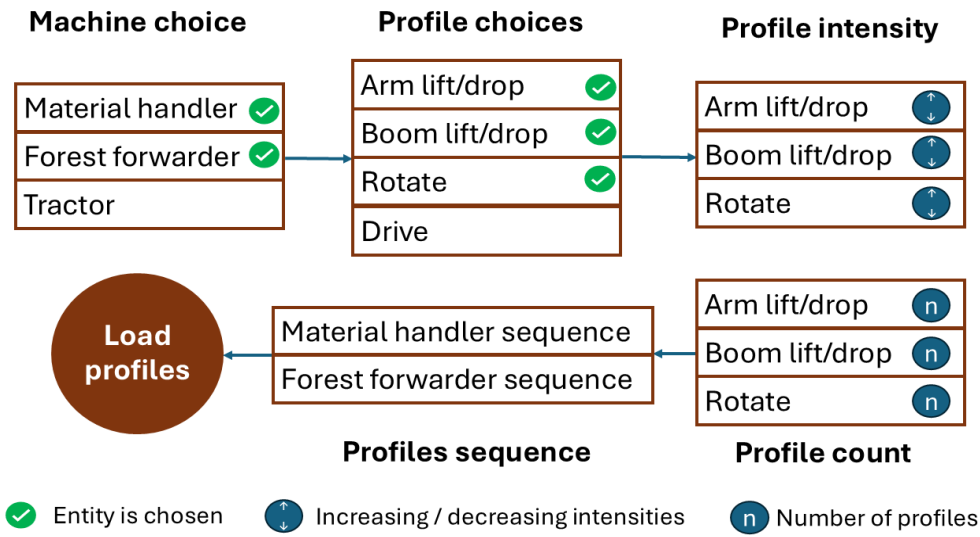

**Figure 9: Example of an aggregation of machinery operational functions.**

This connectivity illustrates the multifunctional nature of the NRMM load profile generator, allowing for tailored machine operations to perform a variety of tasks in different work environments.

## Results

To demonstrate the power requirement of the machines derived by the Load Profile Generator, two case studies for material handler and forest forwarder were conducted. The operating parameters of the selected machines were selected to match the required specifications. To demonstrate the similarity between the power profiles generated by the Load Profile Generator and real-world machinery behavior the validation process was carried out. Some velocities of the hydraulic cylinders, originally taken in m/s, are converted to deg/s to model the angular motion of components such as the arm or boom, considering the geometry of the respective subsystems.

*Case study: Material handler*

The following table (Table 1) shows a working cycle for a material handler where it picks up the load, rotates and drops it at the dump site. The working cycle was constructed by adding the profiles together. The durations were taken from the real-world work cycle of the Atlas 350 MH material handler, whose work cycle was available online [25]. The angular velocity maximum values were assumed based on the duration and acceleration requirements.

Table 1: Example working cycle of the material handler [25].

| Task | Duration [s] | Angular Displacement [deg] | Velocity max [deg/s] |
|---|---|---|---|
| Boom lower | 3 | 70-45 | 15 |
| Arm lift | 3 | 60-90 | 20 |
| Boom lift | 3 | 45-60 | 7.5 |
| Rotate | 5 | 0-50 | 18 |
| Boom lift | 2 | 60-75 | 10 |
| Arm lift | 3 | 90-105 | 10 |
| Rotate | 5 | 50-0 | 18 |
| Arm lower | 4 | 105-60 | 15 |
| Boom lower | 3 | 75-70 | 3 |

To validate the power demand of the material handler, measurement points of the machine working cycle were obtained from a **different** undisclosed (as a part of non-disclosure agreement) company specialized in the material handler equipment manufacturing. Figure 10 shows a side-by-side comparison of the power demand calculations for the boom, arm, and cabin rotation on the right with the chosen section of measurement points on the left of the validation machine, which operated during a similar loading work cycle. The figure demonstrates that the calculated power demand reproduces several of the relative magnitudes found in the measured data, and can therefore can be used for validation.

The main limitation of the validation is that the Load Profile Generator does not mirror the validation machine's measurement points because no kinematic profiles of the machine were available. Therefore, sections where the machine's operating behavior consistently matched the simulation results were identified for validation because the machine repeatedly engaged in similar loading operations. In addition, our model represents a **simplified operational behavior** of the machine, calculated in a sequential manner; therefore, it does not match the level of accuracy of **highly oscillatory** sensor measurements.

The colors in the figure represent the power demand of the same components on both sides. Absolute power numbers were omitted due to confidentiality concerns regarding the measurement points. The validation results showed acceptable comparability between our model and the machine's measured points in some sections of the real-world example, as similar power demand peaks were achieved with the given material handler parameters (e.g., component masses and lengths).

This model also considers the bearing resistance from Liebherr [24] in the calculation of rotation power, resulting in relatively high power peaks. Boom lifting requires more power than arm lifting because the boom-arm system has a greater mass, resulting in a greater maximum force applied to the boom cylinder. The least amount of power is required to lower the boom and arm because the lowering force of the component is negligible, and the hydraulics pressure is assumed to remain at 2.5 bar during lowering.

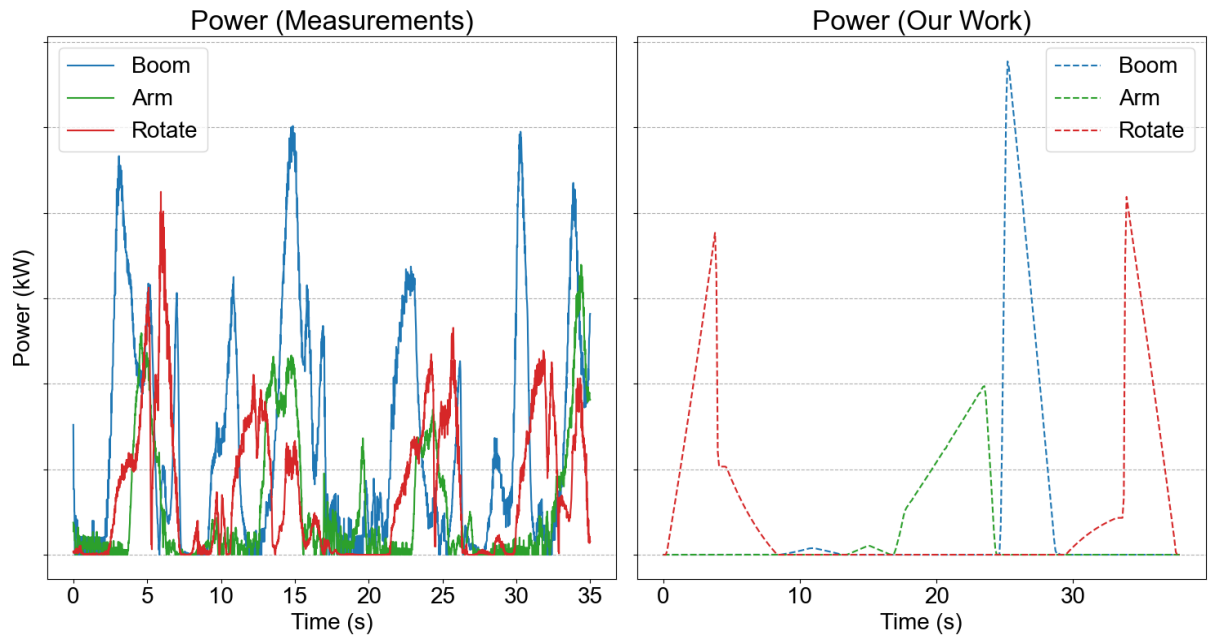

**Figure 10:** Side by side comparison of the bottom-up simulation results for the angular velocity, angular position, and power demand profiles of the selected material handling machine. The validation machine measurements are shown on the left. The results of our work are shown on the right.

*Case study: Forest forwarder*

The inputs for the forest forwarder were observed from the actual machinery operation and represented in the following table (Table 2).

**Table 2:** Chosen operational parameters for the forest forwarder [26].

| Task | Duration [s] | Angular Displacement [deg] | Velocity [deg/s] |
|---|---|---|---|
| Boom lower | 3 | 65-30 | 8 |
| Arm lift | 8 | 90-120 | 16 |
| Boom lift | 3 | 30-65 | 16 |
| Arm lower | 8 | 120-70 | 7 |
| Rotate with load | 4 | 0-90 | 20 |
| Rotate without load | 5 | 90-0 | 20 |

Rotation is modelled separately to complete the machine profile of moving and laying the logs. As can be seen from the simulation results, the maximum power demand occurs during the boom operation due to the boom's bigger weight and an additional weight of the arm.

The power requirements of lifting arm and boom operations are more energy intensive than lowering operations, where the cylinder pressure is assumed to be maintained as low as 2.5 bar, thereby exerting the force on the cylinder walls. The rotation power is also low compared to the boom and arm, because the model does not consider the bearing resistance, but calculates the necessary torque from the moments of inertia, the angles of the components and the acceleration. Also, the cabin of the forwarder was not considered. In absolute terms, the power required to lift the boom and arm is low due to the low mass of the manipulator itself.

For the forest forwarder, the results were compared with the multibody model during the HIL simulation and the virtual simulation of the log-lifting manipulator. Table 3 shows the power demand profile for the manipulator of forest forwarder from Luostarinen [15]. Since the piston speed and power of the lifting cylinder were available, this dataset was used to verify the accuracy of the model. The following parameters were used to model the forest forwarder manipulator and match the parameters in the publications.

**Table 3: Parameters of the forest forwarder manipulator model**

| Parameter | Magnitude |
|---|---|
| $x_c$ / $y_c$ (Eq. 10, Eq. 11) | 0.25 m / 1.5 m |
| $l_{joints(AB)}$ (Figure 8) | 0.5 m |
| $m_{arm}$ / $m_{boom}$ / $m_{load}$ (Figure 8) | 300 / 500 / 260 kg |
| $l_{boom}$ / $l_{arm}$ (Figure 8) | 3.5 / 3.5 m |

Figure 11 shows a side-by-side comparison of our simulation (on the right) and the HIL simulation test bench results (on the left). Validation is achieved by comparing the maximum power peak of the boom, which represents the power that the lower cylinder achieves under the specific boundary conditions outlined in Table 3. The maximum power demand of the lifting cylinder is about 6 kW at the maximum cylinder speed of 1 m/s. Out model shows the maximum boom cylinder power of 5 kW with the adjusted speed of 1 m/s for the boom cylinder. The difference in the results can be explained by the simplicity of our model compared to the HIL, as well as by our estimation of plausible parameters from the sparse available information. Also, the power profile of the multi-body model is smoother than that of the load profile generator due to the controllable acceleration. Similarly, the validation purposes did not aim to exactly replicate the behavior of the logging machinery due to the absence of the kinematic profiles of the validation manipulator. The purpose of validation is to demonstrate the similarity of power profiles with similar parameterization of machinery components.

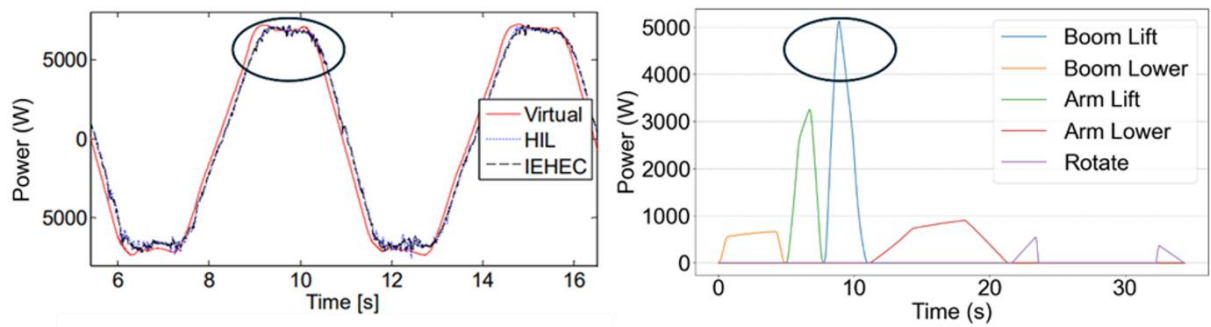

Figure 11: Side by side comparison of the forest forwarder manipulator power demand. The plausibility of the machine was verified by comparing the power demand peaks of the manipulator boom cylinder during the HIL simulation with the control algorithm.

# Conclusion

The development of the Load Profile Generator has demonstrated that a bottom-up modelling approach can plausibly reproduce the operational behavior and model the power demand of diverse machines across the NRMM category. The modelling approach accounts for modularity, reusability and adaptability depending on the operating condition and the machine type. Additionally, validation against real world data confirms the model's capacity to generate plausible load profiles. While the results are plausible, the lack of publicly available data in some areas required the use of assumptions for parameterization and modeling, which may affect precision in specific cases.

The novelty of this work lies in covering about 99% of the machines in the construction, industrial, agricultural and forestry sectors in Germany documented in TREMOD [8]. Machines such as generators, ships, boats, lawn mowers were not included in this study and don't belong to this work's NRMM boundary definitions. Also, the Load Profile Generator can be used to model the power requirements of machines as a whole, as well as those of their individual parts. Due to the flexibility of the tool, it can be further developed for other machine types to expand the pool of machines already presented. This approach is especially important for evaluating powertrain technologies to test their feasibility and optimize machine usage depending on the use case scenarios as well as to identify opportunities to reduce fuel consumption and emissions.

# Acknowledgement

This work was supported by Germany Federal Ministry for Economic Affairs and Energy (BMWE), grant number 03EN5027I.

# Declaration of generative AI and AI-assisted technologies in the manuscript preparation process.

During the preparation of this work the authors used DeepL in order to improve the readability and language of their own texts. After using this tool, the authors reviewed and edited the content as needed and take full responsibility for the content of the published article.

# Appendix

## Symbols and indices

**Table 4: Symbols and indices which are used in this paper**

| Symbols | Description |
|---|---|
| $T[0]$ | Starting time |
| $T[1], T[2], T[3]$ | Time points of the different acceleration phases |
| $v_{max}$ | Used defined input for the maximum velocity |
| $s_{end,calc}$ | Adjusted position to match the velocity |
| $s_{end,set}$ | Used defined input for the end position |
| $a_{ramp,set}$ | Fraction of the accelerating phase |
| $x_{boom}$ | Center of mass of the booms x-axis |
| $L_{boom}$ | Boom length |
| $\theta_1$ | Angle between the boom and the base |
| $\theta_2$ | Angle between the boom and the arm |
| $L_{arm}$ | Arm length |
| $x_{arm}$ | Center of mass of the arms x-axis |
| $x_{bucket}$ | Center of mass of the buckets x-axis |
| $x_{arm,tot}$ | Center of mass of the arm and buckets x-axis |
| $m_{tot}$ | Mass of the complete manipulator system |
| $x_{COM}$ | Center of mass of the complete manipulator systems x-axis |
| $M_{boom}$ | Boom torque |
| $x_c, y_c, z_c$ | Distances between the points where the boom connects to the base |
| $\theta_c$ | The angle to horizontal line from the boom cylinder connection |
| $l_{joints(XY)}$ | Distance of the joints X and Y (refer to the corresponding Figure) |
| $F_{cyl,boom}$ | Force the boom cylinder must exert |
| $D_{cyl,boom}$ | Diameter of the boom cylinder |
| $D_{rod,boom}$ | Diameter of the rod inside boom cylinder |
| $l_{cylinder(XY)}$ | Length of the cylinder XY (refer to the corresponding Figure) |
| $M_{Arm}$ | Torque of the arm |
| $F_{cyl,arm}$ | Force the arm cylinder must exert |
| $m_{grapple + load}$ | Mass of the grapple and the load |
| $\beta$ | Angle between boom and the boom cylinder |
| $\alpha$ | Angle between boom and the vertical line |
| $l_{Arm}$ | Arm length |
| $D_{cyl,arm}$ | Diameter of the arm cylinder |
| $D_{rod,arm}$ | Diameter of the rod inside arm cylinder |
| $v_{cyl,boom}$ | Speed of the boom cylinder |
| $v_{cyl,arm}$ | Speed of the arm cylinder |
| $P_{cyl,boom}$ | Power needed for the boom cylinder |
| $P_{cyl,arm}$ | Power needed for the arm cylinder |
| $I_{cabin}$ | Inertia of the cabin |
| $l_{cabin}$ | Length of the cabin |
| $w_{cabin}$ | Width of the cabin |
| $I_{total}$ | Inertia of the cabin and manipulator system |
| $I_{arm}$ | Inertia of the arm |
| $I_{boom}$ | Inertia of the boom |
| $d$ | Distance between the center of mass of the arm and the axis of rotation |
| $M_{cabin}$ | Torque of the cabin |
| $\omega_{rot}$ | Rotational speed of the cabin |
| $P_{cabin}$ | Power needed for the cabin rotation |

**Plan Routine**

Plan Routine allows the user to select the machine to be simulated and parameterize it according to the scenario. The following code snippet shows an example of how to set up a any machine for simulation.

```
'task_list':[

{'task':'drive',          'params':{'time_ini': a ,'duration': b, 'velocity_max': c, 'ramp %': d
}},
{'task':'boom_lift',      'params':{'time_ini':'append',   'duration':b, 'value_ini': e,
'value_fin': f, 'velocity_max': c, 'load': g}},
{'task':'arm_lift',       'params':{'time_ini':'simultan', 'duration':b, 'value_ini': e,
'value_fin': f,  'velocity_max': c}},

]
```

Listing 1: Task configuration example for a specific machine

The configuration outlined in the code snippet is aimed at setting up a simulation environment for a material handler vehicle. The sequence of setting up a plan for the driving profile is as follows:

1. *Machine Type:* Chose machine type
2. *Defined Task:* Choose task
3. *Task Parameters*:

- *Start Time:* In seconds
- *Duration:* In seconds
- *Start Position:* In meters / degrees
- *End Position:* In meters
- *Velocity:* In m/s

**Machine parameters for the Plan Routine**

<u>Machine parameters</u> contain specific inputs which are fed externally. To execute the above cycle, the material handler model is configured in the Load Profile Generator as follows. For instance, the `'boom_lower'` task initiates the boom lowering action, and `'params':{'time_ini':0,'duration':3,'value_ini':70, 'value_fin':45, 'load': 0, 'velocity_max': 8}` snippet defines the parameters needed to perform this task. The initial time parameter is given in seconds with the specified duration. The tasks can be set to either be appended to start after the previous task, or to be concurrent to start at the same time. The initial and final values specify the initial and final angular positions of the components with respect to their reference axis. The `'velocity_max'` defines the upper limit of the velocity speed, and it is possible to define a ramp or a portion of the acceleration/deceleration time of the total task duration. The load to be handled is defined in the plan, while all other parameters are read separately.

```
'task_list':[

{'task':'boom_lower',       'params':{'time_ini':0,            'duration':3, 'value_ini':  70,
'value_fin':45, 'load': 0, 'velocity_max': 8}},
{'task':'arm_lift_mh',      'params':{'time_ini':'append',     'duration':3, 'value_ini':60,
'value_fin': 90, 'velocity_max':7, 'load': 0, 'boom_fin_ang':45}},
{'task':'boom_lift',        'params':{'time_ini':'append',     'duration':3, 'value_ini': 45,
'value_fin': 60,  'velocity_max':8, 'load': 2000}},
{'task':'rotate',           'params':{'time_ini':'append',     'duration':5, 'value_ini':   0,
'value_fin': 50, 'velocity_max': 6,'ramp %':0.45, 'load': 2000, 'ang_arm': 90, 'ang_boom':60}},
{'task':'boom_lift',        'params':{'time_ini':'append',     'duration':2, 'value_ini': 60,
'value_fin': 75,  'velocity_max':8, 'load': 2000}},
{'task':'arm_lift_mh',      'params':{'time_ini':'append',     'duration':3, 'value_ini':90,
'value_fin': 105, 'velocity_max':7, 'load': 2000, 'boom_fin_ang':75}},
{'task':'rotate',           'params':{'time_ini':'append',     'duration':5, 'value_ini':  50,
'value_fin': 0, 'velocity_max': 6, 'load': 0, 'ramp %':0.45, 'ang_arm': 75, 'ang_boom':75}},
{'task':'arm_lower_mh',     'params':{'time_ini':'append',     'duration':4, 'value_ini':105,
'value_fin': 60, 'velocity_max':7, 'boom_fin_ang':15}},
{'task':'boom_lower',       'params':{'time_ini':'append',     'duration':3, 'value_ini':  75,
'value_fin':45, 'load': 0, 'velocity_max': 8}},

]
```

Listing 2: Code snippet for the material handler's task list in the plan

The task list planning logic remains the same for all machines. The <u>tasks for the forwarder</u> are included in their respective execution order and represented in the Listing 2.

```
task_list = [

{'task': 'boom_lower', 'task_repeats': 1, 'params': {'time_ini': 0, 'duration': 3, 'value_ini':
65, 'value_fin': 30, 'velocity_max': 8, "arm_fin_ang" : 120}},
{'task': 'wait', 'task_repeats': 1, 'params': {'time_ini': 'append', 'duration': 3}},
```

```
{'task':  'arm_lift',   'task_repeats':   1,'params':   {'time_ini':   'append',   'duration':   4,
'value_ini': 90, 'value_fin': 120, 'velocity_max': 7, 'load': 400, 'boom_fin_ang': 30}},
{'task':   'boom_lift',   'task_repeats':   1,'params':   {'time_ini':   'append',   'duration':   3,
'value_ini': 30, 'value_fin': 65, 'velocity_max': 7, 'load': 400, "arm_fin_ang" : 120}},
{'task': 'rotate', 'task_repeats': 1,'params': {'time_ini': 'append', 'duration': 4, 'value_ini':
0, 'value_fin': 90, 'velocity_max': 18, 'ramp %': 0.4, 'ang_arm': 120, 'ang_boom': 65}},
{'task': 'wait', 'task_repeats': 1, 'params': {'time_ini': 'append', 'duration': 3}},
{'task': 'rotate', 'task_repeats': 1,'params': {'time_ini': 'append', 'duration': 4, 'value_ini':
90, 'value_fin': 0, 'velocity_max': 18, 'ramp %': 0.4, 'ang_arm': 120, 'ang_boom': 65}},

]
```

Listing 3: Code snippet for the plan of forest forwarder